\documentstyle[12pt,draft]{article}
\begin{document}

\title{Lattice Formulation of the Standard Model
\thanks{ This
manuscript has been authored under contracts number DE-AC02-76CH00016
and DE-FG02-91ER40676 with the U.S.~Department of Energy.
Accordingly, the U.S.~Government retains a non-exclusive, royalty-free
license to publish or reproduce the published form of this
contribution, or allow others to do so, for U.S.~Government purposes.}
}

\author{Michael Creutz, Michel Tytgat\\
\normalsize
Physics Department,
Brookhaven National Laboratory\\
\normalsize
Upton, NY 11973, USA \\
\\
Claudio Rebbi\\
\normalsize
Department of Physics, 
Boston University\\
\normalsize
Boston, MA 02215,
USA\\
\\
She-Sheng Xue\\
\normalsize
INFN -- Section of Milan,
Via Celoria 16\\
\normalsize
Milan, Italy
}
\date{March 20, 1997}
\maketitle

\begin{abstract}
Combining the Kaplan surface mode approach for chiral fermions with
added terms motivated by Eichten and Preskill suggests the possibility
for a lattice regularization of the standard model which is finite,
exactly gauge invariant, and only has physically desired states in its
low energy spectrum.  The conjectured scheme manifestly requires
anomaly cancelation and explicitly contains baryon and lepton number
violating terms.
\end {abstract}

From the beginnings of lattice gauge theory, chiral symmetries have
been perplexing.  The issues revolve around anomalies and fermion
doubling.  For vector-like theories, such as the strong interactions
via gluon exchange, the problems are largely resolved.  The standard
Wilson \cite{wilson77} approach adds a symmetry breaking term to give
all doublers a mass which becomes infinite with the cutoff scale.  The
approach breaks chiral symmetry rather severely, with the usual
current algebra predictions only expected in the continuum limit.
While somewhat inelegant, the procedure is well defined and widely
adopted.

The situation is more clouded for the full standard model.  Here
chiral symmetry plays a fundamental role, with neutrinos maximally
violating parity.  To couple a gauge field, such as the $W$, to the
requisite chiral currents is considerably less straightforward.  Among
the interesting requirements is the baryon violating process discussed
by 't Hooft \cite {thooft} in the context of topologically non-trivial
gauge configurations.  As emphasized by Eichten and Preskill \cite
{ep} and further discussed by Banks \cite{banks}, a valid lattice
formulation must allow for such processes and incorporate terms which
violate all anomalous symmetries.  Early attempts to include such in
the context of a generalized Wilson action met with difficulties \cite
{antiep}.

A particularly beautiful feature of the original Wilson lattice theory
\cite {wilson74} is its exact local gauge invariance.  While one
possible approach to the standard model is to break chiral symmetry
explicitly, as with the Wilson fermion approach, for the weak
interactions this will also break the gauge symmetry, requiring a
plethora of counter terms \cite {rzref}.  Our goal is a lattice
formulation that keeps all gauged symmetries exact.

A few years ago Kaplan \cite{kaplan} suggested a lattice
generalization of an analysis by Callan and Harvey \cite{callanharvey}
as the basis for a theory of chiral fermions.  The approach uses
topological defects to bind fermionic zero modes\cite{jackiwrebbi}.  A
``domain wall'' in five dimensions can naturally bind chiral states.
In band theory these modes are known as Shockley \cite {shockley}
states, and arise when the particle states and the Dirac sea are
strongly coupled \cite {mcih}.  This approach, and an elegant
variation by Narayanan and Neuberger \cite {nn,nn1,rds}, have
rekindled interest in chiral theories on the lattice.  The extension
to an extra dimension is also quite reminiscent of anomaly effects in
chiral Lagrangian theory \cite{mctytgat}. Nevertheless, subtle
confusion revolves around making the extra dimension infinite
\cite{gjpv,nncritic}.  Here we strive to control this limit, providing
further support for the approach of Refs.~\cite{nn,nn1,rds}.

When the extra dimension is finite, the topological defects are
naturally paired.  For every domain wall there is a mirror defect
carrying additional modes.  This naturally gives rise to a doubling of
species; indeed, this pairing is the minimal amount required by basic
theorems \cite{nielsenninomiya}.  This scheme does provide a promising
approach to the chiral symmetries of the strong
interactions\cite{shamirfurman}, but for the electro-weak theory gives
unwanted ``mirror'' particles.  Here we argue that one can deal
directly with Kaplan fermions on a finite lattice, using a variation
on the Eichten-Preskill idea to give the mirrors masses of the order
of the cutoff.

We start with the standard five dimensional Wilson fermion theory with
hopping parameter sufficiently large that surface modes appear \cite
{mcih}.  Our boundary condition is open in the fifth dimension,
implementing Shamir's \cite {shamir} variation on the Kaplan approach
(this detail is not essential).  We take ordinary space-time
dimensions as periodic.  We add enough fermionic fields to establish
on one four dimensional face of this system all the desired fermionic
states of the standard model, i.e. a strong triplet of weak doublets
of quarks and a lepton doublet for each generation.  We make no
attempt to explain why the real world seems to have three generations,
and thus just repeat this structure three times.  Unlike in
Ref.~\cite{shamirfurman}, we put both the left and right handed
components of the quarks on the same face.  We also include spectator
right handed neutrinos on this wall.  While these decouple in the
standard model, their mirrors are necessary for the removal of other
extraneous states.

At this stage we have the fundamental fermions of the full standard
model on one interface.  However, on the secondary wall in the fifth
dimension an unwanted mirror state exists for each desired mode.  As
usual with the domain wall approach, we couple the four dimensional
gauge fields equally to each slice, and put no gauge field component
in the extra direction.  The mirror states then couple to the gauge
fields with equal strength but opposite parity as the desired
fermions.

We want to give the extra states masses comparable to the cutoff
scale.  We wish to do this without breaking any of the gauge
symmetries.  This problem is mathematically equivalent to eliminating
an extra generation from the standard model; we just have peculiar
parity properties.  To remove a family is inherently non-trivial
because of the 't Hooft process involving baryon decay. The baryon
number change in that process is proportional to the number of
generations; thus, to eliminate one requires additional baryon
violation.

The presence of the 't Hooft process hints at a way to do exactly what
we want.  Indeed, 't Hooft described the process in terms of an
effective interaction vertex.  Considering only a single generation at
the hadronic level, a member of the proton-neutron doublet can convert
to a member of the positron anti-neutrino doublet.  In terms of these
physical particles, such a mixing is an off-diagonal mass term. To
give the particles additional mass, one can artificially enhance this
coupling.  Our suggestion is to add such a coupling only on the
secondary wall, leaving the primary wall bearing all the low energy
fermions of the standard model.  In essence, we use the Kaplan
approach to separate the desired states from their mirrors, and then
apply an Eichten-Preskill interaction to generate a mass gap for the
mirrors.

The weak interactions generate the product over generations of such
vertices only for left handed helicities.  What we do here differs in
two respects.  First, rather than the product of such terms, we treat
the generations independently and add together terms for each.  This
simplifies the discussion so we can treat each family separately.
Second, we add a vertex of this form for each mirror helicity, both
left and right.  This will generate a mass gap for all mirrors.  We
place these terms only on the secondary wall of our five dimensional
formalism.

The above discussion is at the level of the physical particles after
confinement is taken into account.  At another level, the added vertex
is actually a four-fermion coupling, mixing anti-leptons with triplets
of quark fields.  To write the coupling in a compact form, extend the
strong $SU(3)$ index to take on a fourth value representing the
leptons.  We work with a three indexed fermionic field
$\psi_{\alpha,i,s}$ where the first index $\alpha$ represents this
four component combination, the second index represents the two
components rotating under the $SU(2)$ of the weak interactions, and
the final index represents the spinor components of the fermionic
field.  For a chiral fermion, the last index can be restricted to only
two components.  Thus there are a total of 16 independent fermionic
variables for each generation.  Explicitly in terms of the fields for
the $u$ and $d$ quarks and the electron neutrino doublet, these fields
are
\begin{equation}
\Psi=\pmatrix{
u^r_1 & d^r_1 & u^r_2 & d^r_2 \cr
u^g_1 & d^g_1 & u^g_2 & d^g_2 \cr
u^b_1 & d^b_1 & u^b_2 & d^b_2 \cr
\nu_1 & e_1   & \nu_2 & e_2   \cr
}.
\label{matrix}
\end{equation}
Here the subscripts represent the two components of the chiral field,
and the superscripts are the internal symmetry indices of the
quark confining dynamics (QCD). 

The interaction we are interested in is
\begin{equation}
\matrix {V=\epsilon_{\alpha_1\alpha_2\alpha_3\alpha_4}
\epsilon_{i_1i_2}\epsilon_{i_3i_4}
\epsilon_{s_1s_2}\epsilon_{s_3s_4}\cr
\psi_{\alpha_1i_1s_1}\psi_{\alpha_2i_2s_2}\psi_{\alpha_3i_3s_3}
\psi_{\alpha_4i_4s_4}.\cr
}
\label{vertex}
\end{equation}
We add to our Hamiltonian or Lagrangian a tunable coupling $g$ times
$V+V^\dagger$.  Separate vertices are used for left and right handed
fields, although the weak interactions only generate one of these.  We
need both to generate masses for everything on the extra wall.

The invariance of the antisymmetric tensors ensures that this
interaction respects exactly all the desired symmetries of our system.
These include the $SU(3)$ of the strong interactions and the weak
$SU(2)$ symmetry.  The $U(1)$ invariance follows from the neutrality
of the vertex.  The coupling is also a Lorentz scalar since chiral
fermionic fields transform as a rotation by a complex angle, although
as usual this symmetry will be broken by the lattice regularization.

That such a vertex can induce a mass gap follows from a strong
coupling ($g$) expansion about the static limit.  In this limit each
site decouples, giving
\begin{equation}
\int_f \exp \left(g\sum_n (V(n)+V^\dagger(n))\right)=(Cg^8)^N,
\label{static}
\end{equation}
where the $\int_f$ is a path-integral over fermionic fields, $C$ is a
non-zero constant and $N$ is the number of lattice sites on the second
wall.  The power of eight on $g$ comes from sixteen fermion factors
for each of the two helicities, and the vertex is of fourth order.
The kinetic terms for the fermions give a perturbation on this result.

In Hamiltonian language, the basic vertex is a matrix operating on a
Hilbert space of $2^{16}$ basis states.  We normalize with the
conventional anti-commutation relation
\begin{equation}
\left[\psi_{\alpha_1i_1s_1},\psi_{\alpha_2i_2s_2}^\dagger\right]_+
=\delta_{\alpha_1\alpha_2}
\delta_{i_1i_2}\delta_{s_1s_2}.
\label{commutator}
\end{equation}
Regarding the components of $\psi$ as destruction operators and taking
$H=V+V^\dagger$, we have a somewhat unusual quantum mechanics problem,
where fermion number is only conserved modulo four.  

The ground state wave function has fermion number vanishing modulo
four.  It is most easily expressed by applying $V^\dagger$ to the bare
vacuum $|0\rangle$, annihilated by $\psi_{\alpha,i,s}$.  Define the
normalized state $|n> \propto {V^\dagger}^n |0\rangle$.  Because at
most 16 fermions can be created, this sequence terminates at $n=4$.
The Hamiltonian closes on this set, giving a 5 by 5 matrix problem.
The ground state
\begin{equation}
|E_0\rangle ={\sqrt{39}\over 26}\left( |0\rangle+|4\rangle\right)- {1\over 2}
\left(|1\rangle+|3\rangle\right) + {\sqrt {65}\over 13} |2\rangle
\label{groundstate}
\end{equation}
has energy $E_0=-16\sqrt{78} = -141.308\ldots$ and is non-degenerate.
This state is a singlet under both the strong and weak gauge
symmetries.  As expected, it mixes states of different baryon and
lepton number.

Similar manipulations give the first excited state, which turns out to
be in the sector mixing states with fermion number 2 mod 4.  It has
energy $E_1=-8\sqrt {122+10 \sqrt{97}}=-118.79\ldots$.  This energy
represents a multiplet of non-singlet states.

The strong coupling approach starts with each site in the ground
state.  Treated as a perturbation, the fermion kinetic terms allow
hopping between adjacent sites.  This will excite the two sites
involved, requiring a finite energy.  That energy represents a gap in
the spectrum, corresponding to the existence of only massive states.

The enhanced vertex should not induce a spontaneous breaking of one of
the gauge symmetries in the problem.  The unique ground state for the
strong coupling expansion shows that this does not happen as long as
the four-fermion coupling is sufficiently large compared with the
kinetic term.

For our scheme to work, the added coupling must not drastically
interfere with the nature of the heavy states in the fifth dimension.
Ref.~\cite {goltermanshamir} showed such a difficulty with using an
infinite Higgs coupling on the secondary wall.  If we do take our
added coupling to infinity, the last slice in the extra dimension
decouples, giving an effective theory with one less slice.  This
returns us to the starting model with unwanted mirrors.  To avoid this
we must keep the coupling finite but large enough to apply the above
strong coupling analysis on the low energy states.  Ref.~\cite
{goltermanshamir} suggests that a phase with massless mirrors might
persist for a finite range of coupling below infinity.  If that
happens here as well, we must appeal to a hierarchical continuum
limit, adjusting the scale of the extra term to be small compared to
the scale of the heavy states, but large compared with the weak scale.
Here is the weakest point in our argument; a superstrong coupling
phase of massless mirrors and a spontaneously broken region at
intermediate coupling could possibly squeeze out our desired phase of
strongly coupled massive mirrors.  Such a situation would cast serious
doubts on any construction of chiral gauge theories.

A four-fermion vertex can generally be broken into fermion bilinears
interacting with an auxiliary scalar field.  Such Yukawa like models
have been extensively discussed in the past, particularly in the quest
for a chiral fermion theory\cite {yukawa}.  These studies show a
rather rich phase structure.  We want to place the extra wall in a
strong coupling phase with a mass gap but not displaying any
spontaneous symmetry breaking.  Such is sometimes referred to as a
paramagnetic strong coupling phase.
For our purposes we are not interested in a continuum limit of this
Yukawa model; indeed, we want no light particles remaining on the
extra wall as the lattice spacing goes to zero.  Meanwhile, we always
keep the original wall in the weakly coupled phase with light chiral
fermion states.  If the arguments of Ref.~\cite {goltermanshamir} for
a massless phase at ultra-strong coupling hold for our model as well,
then there is a second paramagnetic strong coupling phase which we
must avoid.

Our scheme gives an intuitive description of anomalous currents,
generalizing the discussion in Ref.~\cite{mcih}.  When a topologically
non-trivial gauge configuration induces baryon flow out of the primary
wall of our five dimensional system, this current continues to the
secondary wall where baryon number is strongly violated.  The latter
wall acts as an unusual mirror, reflecting the baryons back as
leptons.  The lepton flow returns to the primary wall and cancels the
lepton number also coming from the topological transition.  Because
there is a mass gap on the secondary wall, it acts as a perfect
mirror, giving no additional factors to the usual tunneling
expression.  In this process the difference of baryon and lepton
number is exactly conserved, just as in the usual continuum standard
model.

Anomaly cancelation is essential to our picture.  In the standard
model both the quarks and the leptons must be present.  Otherwise this
gauge invariant vertex does not exist.  Even though we have strong
baryon violation on one wall, this need not induce unacceptably large
baryon decay for the physical particles.  Communication between the
two domain walls is exponentially suppressed for all but anomalous
currents.  The same small factors as in the usual continuum treatment
\cite{thooft} suppress the latter.

In some sense the theory still has doublers, but we use them.  The
extra term converts the lepton mirrors to composite three
quark states, and the quark doublers to lepton diquark combinations.
Also note that the chiral partner of a given state is a convention
related to the particular gauge field being considered.  From the
standpoint of the strong interactions, the left and right handed
quarks are partners of each other.  For the weak interactions the
matching of particles with doublers is most natural in the above
twisted manner.  At the level of physical particles, the
interpretation is simpler: the doubler of the left handed electron is
the right handed anti-proton.  Both have the same charge, are singlets
under strong $SU(3)$ and are members of weak doublets.

Our proposal makes no use of the Higgs mechanism usually used to
generate particle masses.  Indeed, since our starting point is exactly
gauge invariant, the Higgs generation of physical fermion and weak
boson masses need only be applied in the standard manner at the last
stage.  This might raise new non-perturbative issues, but goes beyond
the subject of this paper.  Since the mirror world has all anomalies
properly canceled, it should decouple from the low energy physics of
the normal standard model in the continuum limit.

Two-dimensional models are often suggested as a testing ground for
chiral theories.  For example, Ref.~\cite{nn96} studies the
electrodynamics of a right moving charge two fermion canceling
anomalies with four left movers of unit charge.  This model introduces
a tricky twist into the scheme we propose.  In particular, the analog
of the 't Hooft vertex contains a derivative.  This requires the
mixing of neighboring sites, and appears to complicate the generation
of the desired mass gap on the secondary wall.  For this case we are
unsure whether a deficiency might lie in our approach.  Indeed, the
extent to which special features of the standard model are crucial to
our approach remains an open question.

While we have used the Kaplan approach to set up the initial model,
presumably one could also work directly with simpler mirror fermion
models\cite{montvay}.  Another direction might be to combine these
ideas with the use of a shift symmetry to separate the doublers as in
\cite{xue}.  Nevertheless, the use of an extra dimension to separate
the problem seems conceptually useful.

If successful, this approach would further justify several alternative
chiral fermion techniques.  For example, the overlap formalism
\cite{nn} effectively represents an infinite extra dimension and
ignores the secondary wall.  Eliminating of that wall in an exactly
gauge invariant manner would support their conclusions as well as
several other schemes that involve an additional infinity \cite
{frolovslavnov,twocutoff,lee}.  Our approach is perhaps cleaner in
that gauge invariance is exact, all infinities are eliminated, and the
requirement of anomaly cancelation is manifest.

\section*{Acknowledgement}  We are grateful for extensive discussions
with M. Golterman and Y. Shamir.

\end{document}